\documentstyle[11pt,moriond,epsfig]{article}

\def\siml{{\ \lower-1.2pt\vbox{\hbox{\rlap{$<$}\lower6pt\vbox{\hbox{$\sim$}}}}\ }}

\def\lQ{\Lambda_{\rm QCD}}

\def\be{\begin{equation}}
\def\ee{\end{equation}}
\def\bea{\begin{eqnarray}}
\def\eea{\end{eqnarray}}

\begin{document}
\vspace*{4cm}
\title{Quarkonium: an effective field theory approach}

\author{Nora Brambilla$^a$  and Antonio Vairo$^b$}
\address{$^a$ Institut f\"ur Theoretische Physik, Universit\"at Wien
Boltzmanngasse 5, A-1090 Vienna, Austria\\
$^b$ Institut f\"ur Hochenergiephysik, \"Osterreichische Akademie der Wissenschaften\\
Nikolsdorfergasse 18, A-1050 Vienna, Austria}

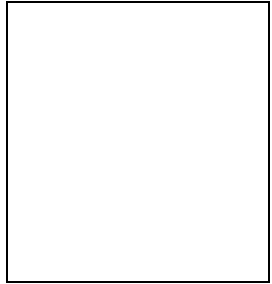
\maketitle\abstracts{We discuss heavy quarkonium in the framework of pNRQCD, a QCD 
effective theory appropriate for the study of heavy quark bound systems.}

\section{Introduction}

The experimental evidence that for heavy quark bound states like $b\bar{b}$, $c\bar{c}$, ... all the 
splittings are considerably less than the masses suggests that all the dynamical energy scales 
of these systems are small with respect to the quark masses. As a consequence the quark velocities $v$ 
are small and these systems can be considered as non relativistic. The hierarchy of the scales is then 
the typical one of a non relativistic system. Called $m$ the mass of the heavy quark, 
the quark momenta scale like $m v$ and the quark energies like $m v^2$.
The inverse of the soft scale, $mv$, gives the size of the 
bound state, while the inverse of the ultrasoft scale, $mv^2$, 
gives the typical time scale. However, in QCD  another physically relevant scale 
has to be considered, namely 
the scale at which nonperturbative effects become important, which we will generically denote by 
$\lQ$.\par
 Trickier is the 
situation in the gluon sector, but the binding interaction is essentially characterized by the 
same energy scale distribution. Therefore the dominant gluon interaction among heavy quarks 
appears ``instantaneous''. A potential picture should hold, at least in first approximation, 
and the energy levels can be obtained by solving the corresponding Schr\"odinger equation.
The study of the heavy quarkonium properties was initiated long ago, using potential models.
A large variety of them exists in the literature and they have been on the
whole phenomenologically quite successful.
However, their connection with the QCD parameters is hidden, the scale at which they are defined  is not
clear, and they cannot be systematically improved. 
In spite of this, great progress has been achieved 
over the years by relating the potentials appearing in these models to some static Wilson loop operators 
\cite{bbp}. 
This formulation is particularly suitable in QCD because it enables a direct lattice calculation 
and/or  an analytic  calculation in a QCD vacuum model \cite{bbp}. However, 
there still remains the 
question of the  extent this pure-potential picture is 
accurate. 

Here, we outline the kind of {\it rigorous description of quarkonium} that  can be obtained 
in the framework of QCD effective field theories.
The dynamical scales of the heavy quark bound system can be  disentangled using effective theories of QCD. From the technical 
point of view, this simplifies considerably the calculation. From the conceptual point of view, this
enables us to factorize the part of the interaction that we know 
from the low energy part which is dominated by nonperturbative physics.

\section{Potential NonRelativistic QCD (pNRQCD)}

The existence of a hierarchy of energy scales $m\gg mv\gg mv^2$, 
allows us to substitute scale by scale QCD with simpler but equivalent effective theories.
Integrating out the hard scale $m$ one passes from QCD to NonRelativisticQCD (NRQCD), while 
integrating out the soft scale $mv$  one passes from NRQCD to potential NRQCD (pNRQCD) 
\cite{pnrqcd}.  
In this last effective theory  only ultrasoft (US) degrees of freedom remain 
dynamical. The surviving fields are the $Q$-$\bar Q$ states  (with US energy) 
and the US gluons.  The $Q$-$\bar Q$ states can be decomposed into a singlet (S) 
and an octet (O) under color transformation. The relative coordinate ${\bf r}= {\bf x}_1-{\bf x}_2$, 
whose typical size is the inverse of the soft scale, is explicit and can be considered as small 
with respect to the remaining (US) dynamical lengths in the system. Hence the gluon fields
can be systematically expanded in $\bf r$ (multipole expansion). Therefore the pNRQCD Lagrangian 
is constructed not only order by order in $1/m$, but also order by order in 
${\bf r}$. As a typical feature of an effective theory, all the non-analytic 
behavior in ${\bf r}$ is encoded in the matching coefficients, which can be interpreted as potential-like terms. 

The equivalence of pNRQCD to NRQCD, and hence to QCD, is enforced to the desired order in the multipole 
expansion by requiring the Green functions of both effective theories to be equal (matching).
The matching between NRQCD and pNRQCD can be done order by order in $1/m$ \cite{pnrqcd}.
Moreover, if we restrict ourselves to the situation $mv \gg \lQ$, we can, in addition, 
do the matching order by order in perturbation theory. 

The  pNRQCD Lagrangian density is given at the next to leading order  in the multipole 
expansion and at order $(1/m)^0$ by:
\begin{eqnarray}
& & \!\!\!\!\!\!\!\!\!\!\!\!\!\! {\cal L}_{\rm pNRQCD} =
{\rm Tr} \Biggl\{ {\rm S}^\dagger \left( i\partial_0  
- V_s(r) + \dots  \right) {\rm S} 
+ {\rm O}^\dagger \left( iD_0 
- V_o(r) + \dots  \right) {\rm O} \Biggr\}
\nonumber\\
& & \!\!\!\!\!\!\!\!\!\!\!\!\!\! \qquad + g V_A ( r) {\rm Tr} \left\{  {\rm O}^\dagger {\bf r} \cdot {\bf E} \,{\rm S}
+ {\rm S}^\dagger {\bf r} \cdot {\bf E} \,{\rm O} \right\} 
  + g {V_B (r) \over 2} {\rm Tr} \left\{  {\rm O}^\dagger {\bf r} \cdot {\bf E} \, {\rm O} 
+ {\rm O}^\dagger {\rm O} {\bf r} \cdot {\bf E}  \right\},  
\label{pnrqcd0}
\end{eqnarray}
where ${\bf R} \equiv ({\bf x}_1+{\bf x}_2)/2$, ${\rm S} = {\rm S}({\bf r},{\bf R},t)$ and 
${\rm O} = {\rm O}({\bf r},{\bf R},t)$ are the singlet and octet wave functions respectively. All the gauge fields in Eq. (\ref {pnrqcd0}) are evaluated 
in ${\bf R}$ and $t$.  In particular ${\bf E} \equiv {\bf E}({\bf R},t)$ and 
$iD_0 {\rm O} \equiv i \partial_0 {\rm O} - g [A_0({\bf R},t),{\rm O}]$. 
$V_s$ and $V_o$ are the singlet and octet matching  potentials respectively;
$V_A$ and $V_B$ are matching coefficients relevant at the next to leading order in the 
multipole expansion. \par
We believe that pNRQCD is the most appropriate effective theory to describe the physics of heavy quark bound systems.
Since pNRQCD 
has potential terms, it embraces potential models. Since in addition it still has ultrasoft gluons 
as dynamical degrees of freedom, it is able to describe non-potential effects.

Indeed, the QED version of it, namely pNRQED, has been shown to correctly reproduce the 
non-potential effects that arise as $O(\alpha^3)$ corrections to the binding energies of hydrogen-like 
atoms and positronium \cite{ps}. Moreover, pNRQCD provides a new interpretation 
of the potentials that appear in the Schr\"odinger equation in terms of a modern effective 
field theory language. The potentials are nothing but {\bf r}-dependent matching coefficients, which appear 
after integrating out scales of the order of the relative momentum in the bound state ($\sim mv\sim 1/r$) or any remaining dynamical scale above $mv^2$. 
Notice that, being matching coefficients, in the particular situation where no physical scales
are between $mv$ and $mv^2$,
 the potentials generically depend on a 
subtraction point $\mu$. But, below the scale 
$mv$ there are still dynamical US  gluons, which are incorporated in the pNRQCD 
Lagrangian.  The dynamics of 
gluons in pNRQCD is cut-off by $\mu$, and this $\mu$ dependence will cancel the explicit $\mu$
dependence of the potential terms when calculating any  physical quantity.
\begin{table}[htb]
\makebox[2cm]{\phantom b}
\begin{tabular}{|c|c|l|l|}
\hline
$mv$&$mv^2$&potential&ultrasoft corrections\\\hline
$\gg \Lambda_{\rm QCD}$&$\gg \Lambda_{\rm QCD}$&perturbative&local condensates\\
$\gg \Lambda_{\rm QCD}$&$\sim \Lambda_{\rm QCD}$&perturbative&non-local condensates\\
$\gg \Lambda_{\rm QCD}$&$\ll \Lambda_{\rm QCD}$&perturbative + & No US (if light quarks\\
&$~$&short-range nonpert.& $\,\,\,$ are not considered)\\ 
\hline
\end{tabular}
\caption{Summary of the different kinematical situations.}
\label{tab2}
\vspace{-0.8cm}
\end{table}
\section{Results}
pNRQCD clarifies several long standing issues. In the case $\lQ < mv$ several situations 
have been examined \cite{pnrqcd} that we summarize in Tab. 1.
Among them we quote:
the proper treatment
of the IR divergences in the singlet potential in perturbative QCD and the determination  of the
the leading log 3-loop contribution to it \cite{pnrqcd}; the construction of the
 suitable effective field theories for the ultrasoft degrees of
freedom when $mv \gg \lQ \gg mv^2$; the model independent study 
of the short-distance limit of the hybrid static potentials.
\par
These results may become relevant in several situations. In particular,
in accurate, model-independent, determinations of the bottom or top mass.
 From the perturbative point of view, the
running of the singlet potential is the first step towards the full
calculation of the leading log correction to the next-to-next-to-leading order
results available at present for the above observables. On the
other hand our analysis provides a model-independent framework to estimate and
parameterize nonperturbative effects.
\par
Finally let us briefly comment upon the situation where $\lQ \sim mv$.
 Most of
the observed charmonium and bottomonium states correspond to this situation, and hence the results presented here cannot be directly applied. 
This situation requires that the matching between NRQCD and pNRQCD is carried out nonperturbatively.
Nevertheless, many of the features observed in this work survive in  a nonperturbative analysis. 
The potential terms can still be obtained in an expansion in $1/m$, 
while the multipole expansion can also be applied for the ultrasoft degrees of
freedom. Moreover, in real QCD the Goldstone bosons (pions and kaons) remain dynamical at
the ultrasoft scale producing non-potential effects. 
Work in this direction is in progress .
\subsection{Acknowledgments}
N. B. acknowledges the TMR contract No. ERBFMBICT961714, A.V. the FWF contract No. P12254.

\end{document}